\documentclass[runningheads]{llncs}

\usepackage{graphicx}
 \usepackage{pdflscape}
 \usepackage{afterpage}
 \usepackage{flafter}
 \usepackage{capt-of}
\usepackage{hyperref}

\usepackage[noadjust]{cite}

\usepackage{hhline}
\usepackage{multicol}
\usepackage{multirow}
\usepackage{amssymb}
\usepackage{amsbsy}
\usepackage{mathtools,amsfonts, enumerate}
 \usepackage{bm}

\usepackage{lipsum}
\usepackage{commath}

\usepackage{tikz}
\usetikzlibrary{arrows,matrix,positioning,shapes,patterns}
\usetikzlibrary{fit}
\usetikzlibrary{calc}
\usetikzlibrary{decorations.pathmorphing,backgrounds,decorations.pathreplacing}
 \usepackage{tikz-3dplot}

\usepackage[textwidth=25mm,textsize=footnotesize,color=red!40]{todonotes}

\usepackage[T1]{fontenc}
\usepackage{textcomp}
\usepackage{listings}
\input{maple-listings-definition.sty} 
\usepackage{enumitem}

 \newcommand{\bth}{\bm{\theta}}
 \newcommand{\bph}{\bm{\phi}}

\begin{document}
\title{Branching out into Structural Identifiability Analysis with Maple: Interactive Exploration of Uncontrolled Linear Time-Invariant 
Structures\thanks{The final authenticated publication is available online at \url{https://doi.org/10.1007/978-3-030-81698-8_27}}}

\titlerunning{Branching out into Structural Identifiability Analysis with Maple}
%
\author{Jason M. Whyte\orcidID{0000-0002-8575-9504}}
\authorrunning{J. M. Whyte}
%
\institute{Australian Research Council Centre of Excellence for Mathematical and Statistical Frontiers (ACEMS), School of Mathematics and Statistics, 
and \\ Centre of Excellence for Biosecurity Risk Analysis (CEBRA), School of BioSciences, University of Melbourne, Parkville, Victoria, Australia \\
\email{jason.whyte@unimelb.edu.au}}
\maketitle              
\begin{abstract}

Suppose we wish to predict a physical system's behaviour. We represent the system by model structure $S$ (a set of related mathematical models defined by parametric relationships between variables), and parameter set $\Theta$. Each parameter vector in $\Theta$ corresponds to a completely specified model in $S$. We use $S$ with system data in estimating the ``true'' (unknown) parameter vector. Inconveniently, $S$ may approximate our data equally well for multiple parameter vectors. If we cannot distinguish between alternatives, we may be unable to use $S$ in decision making. If so, our efforts in data collection and modelling are fruitless.

This outcome occurs when $S$ is not structurally global identifiable (SGI). Fortunately, we can test various structure classes for SGI prior to data collection. A non-SGI result may inform a remedy to the problem. 

We aim to assist SGI testing with suitable Maple 2020 procedures. We consider a class of ``state-space'' structure where a state-variable vector ${\bf x}$ is described by constant-coefficient, ordinary differential equations, and outputs depend linearly on ${\bf x}$. The ``transfer function'' approach is suitable here, and also for the ``compartmental'' subclass (mass is conserved).

Our use of Maple's ``Explore'' permits an interactive consideration of a parent structure, and variants of this produced by user choices. Results of the SGI test may differ for different variants. Our approach may inform the interactive analysis of structures from other classes.
\end{abstract}

\section{Introduction}
Suppose we wish to predict the behaviour of some physical system so that (for example) we can investigate the system's 
response to novel situations. Should we wish to utilise our system knowledge, we would formulate a mathematical model structure
(``structure'' for brevity), say $S$, to represent the system. Broadly speaking, a structure has two main parts. The first is a collection of parametric relationships (e.g. differential equations) relating system features (state variables, ${\bf x}$, which may not be observable), any inputs (or controls, ${\bf u}$), and observable quantities (outputs, ${\bf y}$). The second is a 
parameter space $\Theta$.
Prior to predicting system behaviour with $S$, we must estimate the true parameter vector 
$\boldsymbol{\theta^*} \in \Theta$ from system observations.

Parameter estimation may return multiple (even infinitely-many) equally valid estimates of 
$\boldsymbol{\theta^*}$. Inconveniently, distinct estimates may lead $S$ to produce very different predictions, either for state variables, or for 
outputs beyond the range of our data. In such a case, an inability to distinguish between alternative estimates renders us unable to  confidently use $S$ for prediction. 
Consequently, if we cannot address the question which motivated our study, our efforts in data collection and modelling are unproductive.

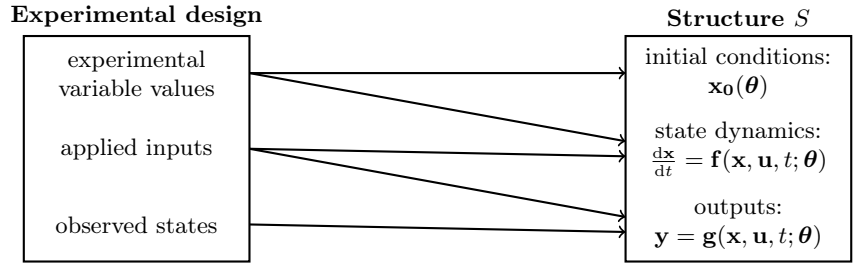
\begin{figure}[hbt] 
\centering
\begin{tikzpicture}[every node/.style={inner xsep=0pt,outer xsep=0pt}]

\node[text width=30mm,minimum height=12mm, minimum width=3cm, align = center] (expdesignmid) at + (0,0){applied inputs};

\node[text width=30mm,minimum height=12mm, minimum width=3cm, align = center, above=0.4cm of expdesignmid.center] (initialval) {experimental \mbox{variable values}};

\node[text width=30mm,minimum height=12mm, minimum width=3cm, align = center, below=0.4cm of expdesignmid.center] (obsstates) {observed states};

\node[draw, minimum width=3cm, minimum height=3cm, black, thick, label={\bf Experimental design}] at (expdesignmid.center) {};

\node[right=5cm of expdesignmid, text width=30mm,minimum height=12mm, minimum width=3cm, align = center] (dynamics) {state dynamics:  \mbox{$ \frac{\dif {{\bf x}}} {\dif t} = {\bf f}({\bf x}, {\bf u}, t; \bth)$}};

\node[above=0.4cm of dynamics.center,text width=30mm,minimum height=12mm, minimum width=3cm, align = center] (ics) {\mbox{initial conditions:} \mbox{${\bf x_{0}}(\bth)$}};

\node[below=0.4cm of dynamics.center, text width=30mm, minimum width=3cm, , minimum height=12mm,  align = center] (output) {outputs: \mbox{${\bf y}= {\bf g}({\bf x}, {\bf u}, t; \bth)$}};

\node[draw, minimum width=3cm, minimum height=3cm, black, thick, label={\bf Structure $S$}] at (dynamics.center) {};

\draw[->,thick] (initialval.east) -- (ics.west) {};

\draw[->,thick,yshift=1mm] (initialval.east) -- ([yshift=1mm]dynamics.west) {};

\draw[->,thick] (expdesignmid.east) -- ([yshift=-1mm]dynamics.west) ;

\draw[->,thick] (expdesignmid.east) -- ([yshift=1mm]output.west) ;

\draw[->,thick] (obsstates.east) -- ([yshift=-1mm]output.west) ;

\end{tikzpicture}
\caption{An illustration of how experimental design in the study of an open-loop system can determine features of a model structure $S$ aiming to represent the system.} \label{fig:Structure-design}
\end{figure}

The problem of non-unique parameter estimates may follow inexorably from the combination of a study design (including planned inputs), and $S$. (To explain further, features of $S$, such as outputs and initial conditions, may follow from the study design. We illustrate this effect for an ``open-loop'' system where outputs do not influence state variables or inputs in Figure~\ref{fig:Structure-design}.)
If so, we can anticipate this problem by testing $S$ subject to its planned inputs for the property of structural global identifiability (SGI). We emphasise that such a test does not require data. Instead, we assume that ``data'' is provided by 
$S$ under idealised conditions. These conditions depend on the class of structure under consideration. However, typical assumptions include:
an infinite, error-free data record is available; and, our structure correctly represents the system. 
When $S$ is an uncontrolled structure, we also assume that the initial state is not an equilibrium state.
Solving algebraic equations derived from $S$ will show whether it is possible (but not certain) for us to obtain a unique estimate of $\boldsymbol{\theta^*}$ under our  idealised conditions. We do not expect a better result for real (noisy, limited) data.

There are other potential rewards for testing $S$ for SGI. Test results may guide the reparameterisation of $S$ into some 
alternative $S^{\prime}$, which may enable parameter estimation to produce a more favourable result than
that achievable for $S$. Similarly, when a structure is not SGI under a given experimental design, one can iteratively examine the potential for alternative designs --- which may produce a modified form of $S$ --- to produce more useful results.

Despite these benefits, the testing of structures for SGI remains uncommon in various domains. This may reflect the specialised nature of identifiability analysis, which requires skills unlike those employed in mathematical modelling. Based on experience, we expect that modellers seeking to scrutinise their model structures will appreciate easy-to-use software tools. We may characterise such tools as those which do not require a somewhat esoteric knowledge of mathematics, or extensive experience with a symbolic algebra package.

We shall use procedures written in the Maple 2020 programming language \cite{Maple2020_v1} to illuminate the testing of structures for SGI.
We demonstrate key concepts through a consideration of continuous-time, uncontrolled, linear time-invariant state-space (henceforth, for brevity, ULTI) structures.\footnote{Broadly, a state-space structure has features as shown for Structure $S$ in 
Figure~\ref{fig:Structure-design}. A ULTI structure includes a collection of linear, constant-coefficient ordinary differential equations that describe the time evolution of state variables.}
More particularly, we consider the ``compartmental'' (that is, subject to
conservation of mass conditions) subclass of ULTI structures, which arise in various modelling applications.
Some standard test methods may not be appropriate for compartmental structures, which guides our choice of test method here.  From an educational standpoint, testing LTI structures for SGI motivates
 the study of various topics, including: systems theory; the Laplace transform; and algebraic equations.

To further extend the value of our procedures, we incorporate these into a routine which automates the testing of a ``parent'' structure for SGI, requiring the user only to define the structure. 
Further, when used with Maple's {\tt Explore}, this routine permits an interactive assessment of the SGI test results obtained for variants of the parent structure (where these variants may be determined by alternative experimental designs).
Experimentation only requires the user to specify (via input boxes) the initial conditions of state variables, and which of these are observed, producing a modified structure and a new application of the SGI test. We are unaware of any other software designed for this purpose.

We also intend to assist those conversant with identifiability analysis. We note recent concerns around reproducibility in computational biology (see, for example, Laubenbacher and Hastings \cite{Laubenbacher_BMB_Editorial_2018}). Reproducibility is impeded when symbolic algebra packages behave inconsistently (as noted for Maple's {\tt assume} command by Armando and Ballarin \cite{Armando_reconstruction_2005}). We intend that our routines will facilitate the checking of SGI test results obtained from either an alternative testing method, or from code written in another language.
We also seek to aid reproducibility with procedures designed to eliminate a source of potential error in structure specification, or to aid the user in recognising other specification errors. This can assist the user in checking  that test results are relevant to the structure of interest.
Additionally, procedures designed for the analysis of LTI structures, possibly with appropriate modification, can assist the testing of linear switching structures (LSSs, which are piecewise LTI) for SGI. (We have explored this previously in the particular context of structures representing biochemical interactions studied on a flow-cell optical biosensor: \cite{jwhyte_On_det_07,MABE07,Whyte_Inferring_2013,Whyte_PhD_2016}.)

The remainder of this chapter is organised as follows.
We present essential definitions pertinent to LTI state-space structures, and 
an outline of concepts useful in testing a (general state-space) structure for SGI in Section~\ref{s:prelims}. We shall focus on the ``transfer function'' (TF) approach --- one of the
earliest methods, yet found in relatively recent textbooks (e.g. \cite{Dynamic_DiStefano_2013}), and one which suits our interest in compartmental structures. Section~\ref{s:Maple_imp} summarises our implementation
of the TF approach in Maple 2020 by outlining our procedures and presenting code listings.
We demonstrate the use of our code and its output by application to a test-case structure in Section~\ref{s:Explore}.
Section~\ref{s:conclusions} offers concluding remarks. 
In \ref{app:figure_code} and \ref{app:Explore_launch} we provide the Maple code used to draw a compartmental diagram, and launch the interactive SGI test,
respectively.

We conclude this section by introducing notation.

\subsection{Notation}

We denote the field of real numbers by $\mathbb{R}$, and its subset containing only positive (non-negative) values 
by $\mathbb{R}_{+}$ ($\bar{\mathbb{R}}_{+}$). The natural numbers $\{1,2,3,\ldots \}$ are denoted by $\mathbb{N}$. The field of complex numbers is denoted by $\mathbb{C}$. Given field $\mathbb{F}$ and some indeterminate $w$, $\mathbb{F}(w)$ denotes the field of
rational functions in $w$ over $\mathbb{F}$. Given $r,c\in \mathbb{N}$ and $\mathbb{F}$, we use $\mathbb{F}^{r \times c}$ to denote the
set of matrices of $r$ rows and $c$ columns having elements in $\mathbb{F}$. 

We use a bold lower-case (upper-case) symbol such as  ${\bf a}$ (${\bf A}$) to denote a vector (matrix), and a superscript ${\top}$ associated with any such object indicates its transpose.  Given vector ${\bf x}$,  $\dot{\bf x}$ denotes its derivative with respect to time.
To specify the $(i,j)$-th element of a matrix, say ${\bf A}$, we may use a lower-case symbol such as $a_{i,j}$, or $({\bf A})_{i,j}$ when this is easier to interpret.
For $n \in \mathbb{N}$, we use ${\bf I}_{n}$ to represent the $n \times n$ identity matrix.

\section{Preliminaries} \label{s:prelims}

In this section we present selected concepts necessary for the development to follow.
We begin in Section~\ref{ss:LTI_structures} by introducing features of ULTI structures.
In Section~\ref{ss:SGI_gen} we provide general definitions for structural global identifiability, and outline 
a process for testing a general state-space structure for this property. We provide details of how to adapt this for ULTI structures in Section~\ref{ss:TF_method}. These details inform the Maple code we shall present subsequently.

\subsection{Linear time-invariant state-space structures} \label{ss:LTI_structures}

LTI state-space structures are appropriate for modelling aspects of various physical applications.
These include quantifying the interconversion of forms of matter in the pyrolysis of
oil-bearing rock (e.g. \cite{Whyte_Woll_book}), or predicting the time evolution of drug concentrations in distinct compartments (say, tissues) of a living subject (e.g. Godfrey~\cite{Godfrey_book}). A key assumption is that the system's state variables (say concentrations) change (e.g. due to metabolic processes, including elimination from the system) according to first-order kinetics (for examples, see Rescigno~\cite{Rescigno_Compartmental_1999}). 

\begin{definition} \label{defn:ULTI}
An {\bf uncontrolled linear time-invariant (ULTI) state-space structure} $M$ with indices $n,k \in \mathbb{N}$ and parameter set $\Theta \subset \mathbb{R}^{p}$ ($p \in \mathbb{N}$) has mappings 
\begin{gather*}
{\bf A}: \Theta \rightarrow \mathbb{R}^{n\times n}, \quad
{\bf C}: \Theta \rightarrow \mathbb{R}^{k\times n}, \quad
{\bf x_{0} }: \Theta \rightarrow \mathbb{R}^{n} \; .
\end{gather*}
The state variables and outputs at any time belong to the ``state space'' $X = \mathbb{R}^n$ and ``output space'' 
$Y = \mathbb{R}^k$, respectively. 
Then, given some unspecified $\bth \in \Theta$, $M$ has ``representative system'' $M(\bth)$ given by 
\begin{align}
\begin{gathered}
\dot{ {\bf x} }(t ; \bth)= {\bf A( \boldsymbol{\theta}) x}(t; \bth) \; , \quad
{\bf x}(0; \bth) = {\bf x_{0}(\boldsymbol{\theta})} \; ,  \\
{\bf y}(t ; \bth) = {\bf C( \boldsymbol{\theta}) x}(t ; \bth) \;  .
\end{gathered}
\label{eq:state_var_rep_uncontrolled} 
\end{align} 
An {\bf uncontrolled positive LTI state-space structure} with indices $n,k \in \mathbb{N}$ is a ULTI state-space structure 
having representative system of the form given in \eqref{eq:state_var_rep_uncontrolled},
where states and outputs are restricted to non-negative values.  That is, the structure has $X = \bar{\mathbb{R}}^n_+$
 and $Y = \bar{\mathbb{R}}^k_+$. 

An {\bf uncontrolled compartmental LTI state-space structure} with indices $n,k \in \mathbb{N}$ is an uncontrolled positive LTI state-space structure composed of systems having system matrices subject to ``conservation of mass'' conditions:
\begin{itemize}
\item all elements of ${\bf C}$ are non-negative, and 
\item for  ${\bf A} =(a_{i,j})_{i,j=1,\ldots,n}$,
\begin{align}
\begin{alignedat}{2}
a_{ij} & \ge  0 \; , & \quad  &  i,j \in \left\{1,\ldots,n \right\}, \quad  i \ne j \; , \\
a_{ii} & \le  - \sum^{n}_{\substack{j=1\\ j \ne i}}  a_{ji} \; , & \quad & i  \in \left\{1,\ldots,n \right\} \; .
\end{alignedat} \label{eq:mass_cons_A}
\end{align}
\end{itemize}
\end{definition}

\subsection{Structural identifiability of uncontrolled structures} \label{ss:SGI_gen}

In their consideration of LTI state-space structures, Bellman and {\AA}ström \cite{Bellman_id} outlined what we may consider as the `classical' approach to testing structures for SGI. Essentially, this involves solving a set of test equations informed by the structure's output, and using the solution set to judge the structure as SGI or otherwise. We pursue this approach following the treatment of ULTI structures in \cite{Whyte_PhD_2016}, which was influenced by Denis-Vidal and Joly-Blanchard  \cite{Denis_auto_equiv04}. 

\begin{definition}[From {\protect Whyte~\cite[Definition 7]{Whyte_MATRIX_IDrev_2020}}] \label{def:ID}
Suppose we have a structure of uncontrolled state-space systems $M$, having parameter set $\Theta$ (an open
subset of $\mathbb{R}^{p}$, $p \in \mathbb{N}$), and time set $T \subseteq [0, \infty)$.
For some unspecified $\bth \in \Theta$, $M$ has representative system $M(\bth)$, which has 
state function ${\bf x}(\cdot; \bth) \in \mathbb{R}^{n}$ and output ${\bf y}(\cdot; \bth) \in \mathbb{R}^{k} $.
Adapting the notation of Figure~\ref{fig:Structure-design} for this uncontrolled case, suppose that the state-variable dynamics and output of system $M(\bth)$ are determined by functions ${\bf f}({\bf x}, \cdot; \bth)$ and ${\bf g}({\bf x}, \cdot; \bth)$, respectively.  
Suppose that $M$ satisfies conditions:
\begin{enumerate}
\item ${\bf f}({\bf x}, \cdot; \bth)$ and ${\bf g}({\bf x}, \cdot; \bth)$  are real and analytic for every $\bth \in \Theta$
on $\mathcal{S}$ (a connected open subset of $\mathbb{R}^{n}$  such that ${\bf x}(t; \bth) \in \mathcal{S}$ for every
$t \in [0, \tau]$, $\tau >0$).
\item ${\bf f}({\bf x_{0}}(\bth), 0; \bth) \neq {\bf 0}$ for almost all $\bth \in \Theta$. 
\end{enumerate}
Then, for some finite time $\tau >0$, we consider the set
\begin{align}
\displaystyle {\mathcal I}(M) \triangleq \left\{ \boldsymbol{\theta^{'}} \in \Theta: 
{\bf y}(t; \boldsymbol{\theta^{'}}) = {\bf y}(t; \bth)  \quad \forall t \in [0,\tau]  \right\} \; . \label{eq:ID_def}
\end{align}
If, for almost all $\bth \in \Theta$:
\begin{description}
\item  $   {\mathcal I} (M) = \{ \bth \}$, $M$ is structurally globally identifiable (SGI);
\item  ${\mathcal I}(M)$ is a countable set,  $M$ is structurally locally identifiable (SLI); 
\item  ${\mathcal I}(M)$ is not a countable set, $M$ is structurally unidentifiable (SU).
\end{description}
\end{definition}

In testing structures from various classes (including the LTI class) for SGI we employ a variant of Definition~\ref{def:ID} that
 is easier to apply. We take advantage of the fact that certain ``invariants'', $\bph(\bth)$, (see Vajda, \cite{Vajda_Structural_1981}),
  completely determine our output function. As such, we may replace the functional equation \eqref{eq:ID_def} with a system of algebraic 
  equations in these invariants.

\begin{definition}[{\protect Whyte~\cite[Definition 8]{Whyte_MATRIX_IDrev_2020}}] \label{def:ID_invariants}
Suppose that structure $M$ satisfies Conditions~1 and 2 of Definition~\ref{def:ID}. Then, for some arbitrary $\bth \in \Theta$, we define
\begin{align}
 {\mathcal I}(M,\bph) \triangleq \left\{ \boldsymbol{\theta^{'} } \in \Theta: 
 \boldsymbol{\phi(\theta^{'})} = \bph(\bth)  \right\} \equiv   {\mathcal I}(M) \; ,
\label{eq:ID_def2}
\end{align}
and determination of this allows classification of $M$ according to Definition \ref{def:ID}.
\end{definition}

\begin{remark}
In the analysis of (say, uncontrolled) LSS structures, there are some subtleties to Definition~\ref{def:ID_invariants}.
It is appropriate to consider the response on independent time intervals between switching events as the same parameter vector does not apply across all such intervals. It is appropriate to re-conceptualise invariants as a collection of features across the time domain; each interval between switching events contributes features which define the structure's output on that interval (\cite{jwhyte_On_det_07,MABE07}). 
\end{remark}

When Definition~\ref{def:ID_invariants} is appropriate for the class of structure at hand, we may employ this at the end of 
a well-defined process, which we summarise below. 

\begin{proposition}[A general algorithm for testing a structure for SGI, from {\protect Whyte~\cite[Proposition~1]{Whyte_MATRIX_IDrev_2020}}] \label{prop:ID_test}
\begin{description}
\item[] Given some model structure $M$ with parameter set $\Theta$, having representative system $M(\bth)$ for 
unspecified $\bth \in \Theta$:
\item[Step 1] Obtain invariants $\bph(\bth)$:
there are various approaches, some having conditions (e.g. that $M$ is structurally minimal, see Remark~\ref{rem:gen_min}) that may be difficult 
to check. 
\item[Step 2] Form alternative invariants $\bph(\bth^{\prime})$ by substituting $\bth^{\prime}$ for $\bth$
in $\bph(\bth)$.
\item[Step 3] Form equations $\bph(\bth^{\prime})= \bph(\bth)$.
\item[Step 4] Solve these equations to obtain $\bth^{\prime} \in \Theta$ in terms of $\bth$ to determine ${\cal I}(M,\bph)$.
\item[Step 5] Scrutinise ${\cal I}(M,\bph)$ so as to judge $M$ according to Definition~\ref{def:ID_invariants}.
\end{description}
\end{proposition}

The particularities of Proposition~\ref{prop:ID_test} depend on both the class of the structure under investigation, and the testing method
we will employ. In the next subsection we provide an overview of the TF method, which is appropriate for the compartmental LTI 
structures of interest to us here.

\subsection{The transfer function method of testing uncontrolled LTI state-space structures for SGI} \label{ss:TF_method}

The TF method makes use of the Laplace transform of a structure's output function (causing an alternative name, e.g. \cite{Godfrey_book}). As such, it is appropriate to recall the Laplace transform of a real-valued function.
\begin{definition}
Suppose some real-valued function $f$ is defined for all non-negative time. (That is,
$f: \bar{\mathbb{R}}_{+} \mapsto \mathbb{R} , \  t  \mapsto f(t)$.)
We represent the (unilateral) Laplace transform of $f$ with respect to the transform
variable $s \in \mathbb{C}$ by
\begin{equation*}
\mathcal{L} \{ f \} (s) \triangleq \int_{0}^{\infty} f(t)\cdot e^{-st} {\rm d}t \; ,
\end{equation*}
if this exists on some domain of convergence $\mathcal{D} \subset \mathbb{C}$.
\end{definition}

When applying the TF to the output of a controlled LTI structure, we must check to ensure that $\mathcal{D}$ exists. However, 
given an ULTI structure having finitely-valued parameters (a physically realistic assumption), each component of ${\bf x}$ 
or ${\bf y}$ is a sum of exponentials with finite exponents which depend linearly on $t$. As such, the Laplace transform does exist on some domain of convergence, the specific nature of which is unimportant for our purposes here. (We direct the reader interested in details to Sections~2.3.1 and 3.1 of Whyte~\cite{Whyte_MATRIX_IDrev_2020}.)

Given ULTI structure $S$ having representative system $S(\bth)$ informed by ${\bf A}(\bth) \in \mathbb{R}^{n \times n}$ and ${\bf C}(\bth) \in \mathbb{R}^{k \times n}$, we may write the Laplace transform of the output function of $S(\bth)$ as:
\begin{align}
\mathcal{L} \{ { \bf y} (\cdot ; \bth ) \}(s;\bth) & = { \bf H_{2}}(s; \bth) \; , \label{eq:LT_uncontrolled_y} 
\end{align}
where \eqref{eq:LT_uncontrolled_y} exists on domain of convergence $\mathcal{C}_{0}$, and the ``transfer matrix''
is\footnote{We have adapted the notation of \cite[Chapter~2]{Walter_book} to include ${\bf x_{0}}$, as
otherwise initial-condition parameters do not appear in the SGI test.}
\begin{align}
 { \bf H_{2}} (s; \bth) & \triangleq {\bf C}(\bth) \Big( s {\bf I}_{n} - {\bf A}(\bth) \Big)^{-1} { \bf x_{0}} (\bth)
 \in \mathbb{R}(s)^{k \times 1} \;  . \label{eq:H2}
\end{align}

The elements of ${\bf H_{2}}$ (``transfer functions'') are rational functions in $s$. We refer to these functions as
``unprocessed'' if we have not attempted certain actions. We must undertake one or more of these in order 
to obtain invariants from ${\bf H_{2}}$ for testing $S$ for SGI. We shall describe these steps and their result for the case of compartmental ULTI structures in the following definition.

\begin{definition}[Canonical form of a transfer function (adapted from {\protect\cite[Definition~9]{Whyte_MATRIX_IDrev_2020}})] 
 \label{defn:canonical_form}
Given compartmental ULTI structure $S$ of $n \in \mathbb{N}$ states, suppose that associated with $S(\bth)$ 
is a transfer matrix ${\bf H_{2}}$ (as in \eqref{eq:H2}), composed of unprocessed transfer functions. (Recall that we know $\mathcal{L}\{ {\bf y } \}$
exists on some domain $\mathcal{C}_{0} \subset \mathbb{C}$, and hence that ${\bf H_{2}}$ is defined.)
Given element $\left({\bf H_{2}}(s;\bth)\right)_{i,j} \in \mathbb{C}(s)$, we must cancel any common factors between the numerator and denominator polynomials (``pole-zero cancellation''). Following this, we may
choose to obtain the associated {\it transfer function in ``canonical form''} by rewriting it to ensure that the denominator is monic.
The result is an expression of the form:
\begin{gather} 
 \begin{gathered}
\left({\bf H_{2}}(s;\bth)\right)_{i,j}  = 
 \frac{\omega_{i,j,r+p}(\bth) s^{p} + \cdots + \omega_{i,j,r}(\bth) }{s^{r} + \omega_{i,j,r-1}(\bth) s^{r-1}+ \cdots + 
  \omega_{i,j,0}(\bth)}, \quad \forall s \in \mathcal{C}_0  \; ,  \\
 r \in \{ 1, \dots, n\} \; , \quad   p \in \{ 0, \dots,  r-1\} \; .
 \end{gathered} \label{eq:LT_output}
\end{gather}
The coefficients $\omega_{i,j,0}, \ldots, \omega_{i,j,r+p}$ in \eqref{eq:LT_output} contribute invariants towards $\bph(\bth)$. 

We may prefer to retain a non-monic denominator if this is desirable, such as when coefficients of $s$ are polynomial in 
$\bth$, but would not be if the denominator was rewritten to become monic. Given a non-monic denominator, we obtain all coefficients of the transfer function to use as invariants. In our procedures we give the user choice on whether or not to require that transfer functions have monic denominators.
\end{definition}

\begin{remark} \label{rem:gen_min}
Various approaches to testing an LTI structure $S$ for SGI (e.g. the similarity transform method) are only 
applicable to a ``structurally minimal'' $S$. 
We see that $S$ is not structurally minimal if we can reduce it to a structure $\tilde{S}$ of $n_{1} < n$ state variables (and, say, parameter set 
$\tilde{\Theta}$) where, for almost all $\bth \in \Theta$, there is some $\tilde{\bth} \in \tilde{\Theta}$ such that the outputs of $S(\bth)$ and $\tilde{S}(\tilde{\bth})$ are identical.
The TF method has the advantage of not requiring structural minimality. 
Instead, undertaking any possible pole-zero cancellation in transfer functions (as required by Definition~\ref{defn:canonical_form}) allows the test to access the parameter information available in a structurally minimal form of $S$.

In the testing of an uncontrolled LSS structure for SGI using procedures presented here, checking for pole-zero cancellation
 in the constituent LTI structures in effect after the first switching event is typically not trivial. This has led to indirect (\cite{Whyte_Inferring_2013}) and direct (\cite{Whyte_PhD_2016}) approaches involving far greater algebraic complexity. 
\end{remark}

In the next section we present the Maple procedures we shall use in testing a ULTI structure for SGI. The source code is available
for download from \cite{Whyte_Maple2020_supportingCode_2021}.

\section{An implementation of the Transfer Function method for uncontrolled LTI state-space structures} \label{s:Maple_imp} 
In Section~\ref{ss:components} we show our procedures for an implementation of the TF method in order of use (according to a general scheme such as Proposition~\ref{prop:ID_test}), and explain certain key features in our specific context. In Section~\ref{ss:proc_integration} we combine these component procedures into a complete SGI test procedure. We validated our
procedures by applying them to structures scrutinised in the literature (e.g. two variants of a three-compartment LTI structure in DiStefano \cite[Example~10.3]{Dynamic_DiStefano_2013}), and confirming that our results were equivalent. 

\subsection{Component procedures} \label{ss:components}
Procedures  {\tt process\_matrix} (Listing~\ref{code:process_matrix}), {\tt collect\_invariants} (Listing~\ref{code:collect_invariants}) 
and {\tt identifiability\_eqn\_list} (Listing~\ref{code:id_eqns}) were adapted from 
Maple 2015 (\cite{Maple2015_v2}) routines presented in Whyte~\cite[Appendix~B]{Whyte_PhD_2016}. Here we have updated those original routines for Maple 2020 \cite{Maple2020_v1}. We have also taken steps to make the original routines more efficient and concise, such as by replacing some loops with {\tt map} commands, or using more appropriate types of data structures. Further, we have improved upon 
{\tt process\_matrix}; previously the routine merely flagged a non-monic denominator in a transfer function. The revised procedure 
uses the logical parameter {\tt canonical\_form}, which specifies whether or not transfer function denominators should be made monic.
This choice may influence the procedure's output:  a processed transfer function matrix. As this matrix is passed to 
{\tt collect\_invariants}, we have adapted this procedure accordingly.

\begin{remark}
Aside from its role in the `classical' approach to testing structures for SGI, historically there was another reason to ensure that each transfer function had a monic denominator. This step enabled the comparison of elements of transfer matrices drawn from two different structures. If each pair of rational functions have the same coefficients, then the two structures produce the same output. (Finding the parameter vectors which cause this equality relates to whether or not two structures have the property of ``structural 
indistinguishability'', a generalisation of structural identifiability; we test for the former using methods similar to those used for the latter.) However, given the symbolic algebra packages available now, monic denominators are not essential. Our code permits the user to allow non-monic denominators in transfer functions by setting {\tt canonical\_form:=false}.
\end{remark}

 Procedure {\tt process\_matrix} (Listing~\ref{code:process_matrix}, the start of Step 1 of Proposition~\ref{prop:ID_test} in this setting) prepares the transfer matrix  associated with a structure $S$ ({\tt transfer\_matrix}) for the extraction of invariants. (Recall the
discussion in Section~\ref{ss:TF_method}.) The {\tt sort\_order} list parameter directs {\tt sort} in how to
order parameters and the complex variable (say $s$) which appear in the processed transfer functions. For each of these, the procedure returns the numerator and denominator polynomials, stored in a matrix.

{ \scriptsize
\begin{lstlisting}[caption={Procedure {\tt process\_matrix} for processing a matrix of transfer functions obtained from a LTI structure to enable the subsequent extraction of invariants.},label=code:process_matrix, language=maple]
process_matrix:=proc(sort_order::list(symbol), transfer_matrix::Matrix, canonical_form::truefalse, s::symbol:="s", $)
local i:=0, j:=0, colMAX, rowMAX, current_element, leading_denom_coeff, processed_matrix, new_numer, new_denom;
description "Prepare matrices of transfer functions for extraction of invariants. This involves pole-zero cancellation, and conversion of transfer functions into their canonical form (by ensuring denominators are monic) if ''canonical_form'' is set to true."; 
rowMAX, colMAX:=LinearAlgebra[Dimensions](transfer_matrix);
processed_matrix:=Matrix(rowMAX, colMAX);
for i to rowMAX do;
for j to colMAX do;
current_element:=normal(transfer_matrix[i, j]);
leading_denom_coeff:=lcoeff(denom(current_element), s);
if (canonical_form=true and leading_denom_coeff<>1) then new_numer:=eval(numer(current_element)/leading_denom_coeff);
new_denom:=eval(denom(current_element)/leading_denom_coeff);
else new_numer:=numer(current_element);
new_denom:=denom(current_element);
fi;
new_numer:=sort(collect(new_numer, s), sort_order, plex);
new_denom:=sort(collect(new_denom, s), sort_order, plex);
processed_matrix[i, j]:=[new_numer, new_denom];
od;
od;
return processed_matrix;
end proc;
\end{lstlisting}
}

Procedure {\tt collect\_invariants} (Listing~\ref{code:collect_invariants}, the conclusion of Step 1 of Proposition~\ref{prop:ID_test}) extracts the coefficients from a processed transfer matrix, including the invariants. (Later in Listing~\ref{code:Explore_SGI_routine} we process the returned object to disregard any numeric coefficients.)

{ \scriptsize
\begin{lstlisting}[caption={Procedure {\tt collect\_invariants} which extracts the invariants from a processed transfer matrix.}, label=code:collect_invariants, language=maple]
collect_invariants:=proc(processed_matrix::Matrix, s::symbol, $)
local i:=0, j:=0, colMAX, rowMAX, latest, coeff_set:={}, element_numer_coeffs, element_denom_coeffs;
description "Extract the invariants from a matrix of transfer functions placed in the canonical form.";
rowMAX, colMAX:=LinearAlgebra[Dimensions](processed_matrix);
for i from 1 to rowMAX do;
for j from 1 to colMAX do;
latest:={};
element_numer_coeffs:={coeffs(processed_matrix[i,j][1],s)};
element_denom_coeffs:={coeffs(processed_matrix[i,j][2],s)};
latest:=element_numer_coeffs union element_denom_coeffs;
coeff_set:=coeff_set union latest;
od;
od;
return map(primpart,coeff_set);
end proc;
\end{lstlisting}
}

 The procedure {\tt theta\_prime\_creation} (Listing~\ref{code:theta_prime}, the start of Step 2 of Proposition~\ref{prop:ID_test}) is new. This routine intends to remove a point in SGI analysis at 
 which human error could cause a mismatch between the ordering of parameters in $\bth$ and $\bth^{\prime}$, potentially causing an
inaccurate test result. The list of the structure's parameters {\tt theta} is modified to return the alternative parameter list {\tt theta\_prime}. This process ensures that there is a clear relationship between corresponding elements of $\bth$ and $\bth^{\prime}$ (to aid interpretation of \eqref{eq:ID_def2}), and the correspondences are correct. 
When {\tt theta\_mod\_type} equals ``underscore'', an element of {\tt theta\_prime} is defined by adding an underscore suffix to the 
 corresponding {\tt theta} element  (line 6). Alternatively,  when {\tt theta\_mod\_type} equals ``Caps'' {\tt theta\_prime} is populated by 
capitalised versions of {\tt theta} (line 7). This option is appropriate when {\tt theta} only contains entries which begin with a lower-case alphabetic character.
 
 { \scriptsize
\begin{lstlisting}[caption={Procedure {\tt theta\_prime\_creation} creates a recognisable alternative parameter from each element of the original parameter vector $\bth$.},label=code:theta_prime, language=maple]
theta_prime_creation:=proc(theta::list(symbol), theta_mod_type::identical("underscore","Caps"):="Caps", $)::list(symbol);
local i, theta_prime, common_params;
description "A list of symbols theta is modified to create list theta_prime such that the connection between original and alternative parameters is apparent. For theta_mod_type=''Caps'' (the default), modify to upper case. For theta_mod_type=''underscore'', append an underscore (_) to each symbol. To specify a subscripted parameter in theta, use two underscores, e.g. k__1, not k[1].";
theta_prime:=Array(theta);
for i from 1 to numelems(theta) do;
if (theta_mod_type="underscore") then theta_prime[i]:=convert(StringTools[Insert](theta[i],length(theta[i]),"_"),symbol);
elif (theta_mod_type="Caps") then theta_prime[i]:=convert(StringTools[UpperCase](theta[i]), symbol); fi;
od;  
# Check that the use of theta_mod_type has created elements of theta_prime which differ from all elements of theta. (For example, using theta_mod_type="Caps", if theta contained upper-case symbols, then, inappropriately, theta_prime would also contain these.)
common_params:=convert(theta,set) intersect convert(theta_prime,set);
if (numelems(common_params) >0) then error "Inappropriate theta parameter(s) for nominated theta_mod_type:", common_params fi;
return convert(theta_prime,list);
end proc;
\end{lstlisting} 
}

Procedure {\tt identifiability\_eqn\_list} (Listing~\ref{code:id_eqns}, concluding Step 2 and Step 3 of Proposition~\ref{prop:ID_test}) uses the structure's invariants $\bph(\bth)$, and parameter vectors $\bth$ and $\bth^{\prime}$, and
returns the necessary SGI test equations $\bph(\bth)=\bph(\bth^{\prime})$.
{ \scriptsize
\begin{lstlisting}[caption={Procedure {\tt identifiability\_eqn\_list} forms the SGI test equations.}, label=code:id_eqns, language=maple]
identifiability_eqn_list:=proc(invariants::list, theta::list(symbol), theta_prime::list(symbol), $)
description "Use a structure's invariants in forming equations necessary for the structural global identifiability test.";
return invariants =~ subs(theta =~ theta_prime, invariants);
end proc;
\end{lstlisting}
}

Procedure {\tt classify\_solutions} (Listing~\ref{code:classify}) is also new. It addresses Step~5 of Proposition~\ref{prop:ID_test} by scrutinising the solutions of the SGI test equations.
{ \scriptsize
\begin{lstlisting}[caption={Procedure {\tt classify\_solutions} classifies the structure as SGI, SLI, SU, or "unknown" if classification is not possible.}, label=code:classify, language=maple]
classify_solutions:= proc(solset::list, theta::list(symbol), theta_prime::list(symbol), $)
local num_soln_families:=numelems(solset), i, lhsides, rhsides, free_param_by_soln:=Vector(num_soln_families), classification:=["Unknown: inspect solutions",magenta], difference, RootOf_check, RootOf_count;
description "Use the solution set of the SGI test equations in classifying the structure under investigation.";
# Initially the structure is unclassifed. It is SGI if there is exactly one solution family, and it is theta_prime equals theta.
if (num_soln_families =1 and verify(solset[1],theta_prime=~theta)=true) then classification:=["SGI",green]; return (classification); fi;
# The structure is SU if any of the solution families contain free parameters. If so, the difference of the left and right-hand-sides of an equation in solset is zero. Look for free parameters over each solution family in solset in turn. As soon as we find a free parameter, classify the structure as SU and exit the routine.
for i from 1 to num_soln_families do;
lhsides := map(lhs, solset[i]); 
rhsides := map(rhs, solset[i]); 
difference := simplify(lhsides - rhsides); 
free_param_by_soln[i] := select(x -> x = 0, difference); 
if (numelems(free_param_by_soln[i])>0) then classification:=["SU",red]; return (classification); fi;
od;
# If the flow has continued to this point, any solution family in solset does not have free parameters (not SU) and does not have a unique solution (not SGI). If there are multiple families, or a family contains roots, the structure is SLI.
if (num_soln_families > 1) then classification:= ["SLI (multiple solutions)",yellow]; return (classification); fi; 
# If we reach this point, we have a single solution family. Does it contain roots?
rhsides := map(rhs, solset);
RootOf_check:= map(type,rhsides,RootOf);
RootOf_count:= select(x -> x = true, RootOf_check); 
if (RootOf_count >0) then classification:=["SLI (solution contains 'RootOf')",yellow]; return (classification); fi;
# If we reach this point, we have not classified the structure.
return (classification);
end proc;
\end{lstlisting}
}

\begin{remark}
The procedures {\tt classify\_solutions}, {\tt identifiability\_eqn\_list}, and  {\tt theta\_prime\_creation}  are not restricted to use in testing LTI structures for SGI. Also, each of the procedures in this section may be used in testing a controlled LTI state-space structure for SGI.
\end{remark}

In the next subsection we combine our component procedures into a complete procedure for testing a ULTI state-space structure for SGI. 
Subsequent use of this with {\tt Explore} allows us to interactively test a parent structure and its variants.

\subsection{A complete SGI test procedure for ULTI state-space structures} \label{ss:proc_integration}

Given some defined structure, Listing~\ref{code:Explore_SGI_routine} forms the transfer matrix ${\bf H_{2}}(s; \bth)$, then draws on Listings~\ref{code:process_matrix} to \ref{code:classify} in applying steps of the SGI test.
We call our procedure {\tt Uncontrolled\_Lin\_Comp\_Fig} (\ref{app:figure_code}) to draw a (modified) compartmental 
diagram associated with the structure as part of the output, which also shows $\bth$,  $\bth^{\prime}$, the solution set of the SGI test equations \eqref{eq:ID_def2}, and a classification of the structure. Use of the procedure via Listing~\ref{code:Explore_application} 
produces text-input boxes which permit the user to modify values of observation gains (elements of ${\bf C}$) and initial conditions. 
Also, drop-down menus permit the user to select the value of {\tt theta\_mod\_type} used in creating $\bth^{\prime}$, or the select the diagram layout style from the options provided by {\tt DrawGraph}.

{ \scriptsize
\begin{lstlisting}[caption={{\tt Explore\_SGI\_test} combines routines from Section~\ref{ss:components}
resulting in a procedure suitable for testing an ULTI structure for SGI.}, label=code:Explore_SGI_routine, language=maple]
Explore_SGI_test:=proc(A::Matrix, obs_gains::list, ICs::list, outflow_params::list, layout_style, theta_mod_type::identical("underscore","Caps"), s::symbol, canonical_form::truefalse, tracing::truefalse, $)
local C, n, x0, x_colour, y_colour, outgraph, G, prelim1, H2, H2_proc, theta, sort_order, coeff_collection, phi_list, i, eqn_list, phi1vec_list, new_list, theta_prime, solset, classification, solset_Matrix, textmatrix, textplot1, textplot2, textplot3;
description "A procedure (for use with Explore) to allow interactive testing of a parent (continuous time, uncontrolled, linear, time-invariant state-space structure and variants derived from it. The user creates variants by setting the structure's initial conditions (ICs) or observation gains (obs_gains) to constants, including zero. The output includes the original parameters theta and alternative parameters theta_prime, the solution set of the test equations, and a type of compartmental diagram showing the dependencies between outputs and state variables.";
interface(rtablesize = 15);
C:=LinearAlgebra[DiagonalMatrix](obs_gains);
n:=LinearAlgebra[RowDimension](A);
x0:=Vector[column](ICs);
x_colour:="LightBlue";
y_colour:="LightGreen";
outgraph:=Uncontrolled_Lin_Comp_Fig(A, C, x0, x_colour, y_colour, outflow_params, tracing);
G:=GraphTheory[DrawGraph](outgraph, layout=layout_style, stylesheet=[vertexhighlightborder=false, vertexborder=false]);
prelim1:=LinearAlgebra[MatrixAdd](LinearAlgebra[IdentityMatrix](n), A, s, -1);
H2:=LinearAlgebra[Multiply](C, LinearAlgebra[Multiply](LinearAlgebra[MatrixInverse](prelim1), x0));
# Ensure that we have a transfer matrix, not a scalar (as could happen when C is a row vector) or a vector
if (type(H2,`+`)=true) then H2:=convert([H2], Matrix); else
H2:=ArrayTools[Reshape](H2, [n, 1]); fi;
theta:=convert(`union`(indets(A), indets(x0), indets(C)), list);
sort_order:=[s, op(theta)];
H2_proc:=process_matrix(sort_order,H2,canonical_form,s);
coeff_collection:=convert(collect_invariants(H2_proc,s),list);
# Here retain only the coefficients that depend on system parameters: exclude elements that are type numeric.
phi_list:=remove(type, coeff_collection,numeric);
theta_prime:=theta_prime_creation(theta,theta_mod_type);
eqn_list:=identifiability_eqn_list(phi_list,theta,theta_prime); 
solset:=solve(eqn_list,theta_prime);
classification := classify_solutions(solset, theta, theta_prime);
solset_Matrix:=convert(solset,Matrix);
# Display the outputs as an array of objects. 
textmatrix:=Matrix([["theta",op(theta)],["theta_prime",op(theta_prime)]]):
textplot1:=plots[textplot]([0,0,textmatrix],axes=none):
textplot2:=plots[textplot]([0,0,solset_Matrix],axes=none,title="Solutions for theta_prime in terms of theta"):
textplot3:=plots[textplot]([0,0,classification[1],'font'=["times","roman",20]],axes=none,title="Structure classification:",background=classification[2]):
plots[display](Array(1..4,1..1, [[textplot1],[textplot2],[G],[textplot3]]));
end proc;
\end{lstlisting}
}

\section{Towards interactive inspection of the effect of changing experimental designs on the SGI test}
\label{s:Explore}

We consider a parent compartmental ULTI state-space structure (as in Definition~\ref{defn:ULTI}) of three compartments, as we may find in pharmacological applications. We assume that we can observe each state variable. We may obtain simpler variants of the structure (reflecting changes to the experimental design, but not the nature of the physical system) by setting any parameter in ${\bf x_{0}}$ or ${\bf C}$ to a non-negative constant. We employ notation for parameters in ${\bf A}$ (rate constants) common to pharmacological applications: $k_{ij}$, ($i \neq j$, $j \neq 0$) relates to the flow of mass from~$x_{j}$ to $x_{i}$, and $k_{0j}$ relates to the outflow of mass from $x_{j}$ to the environment (see Godfrey~\cite[Chapter~1]{Godfrey_book}.)

We specify the structure by:
\setcounter{MaxMatrixCols}{11} 
\begin{align}
\begin{gathered}
{\bf x}(\cdot ;\bth ) =   \begin{bmatrix} x_{1}(\cdot;\bth) \\ x_{2}(\cdot;\bth) \\ x_{3}(\cdot;\bth) \end{bmatrix}\; ,  \qquad
{\bf x_{0}}(\bth) =   \begin{bmatrix} x_{0_1}\\ x_{0_2} \\ x_{0_3} \end{bmatrix}\; ,  \qquad
{\bf y}(\cdot ; \bth) = \begin{bmatrix} y_{1}(\cdot;\bth) \\ y_{2}(\cdot;\bth) \\ y_{3}(\cdot;\bth) \end{bmatrix} \; , \\
{\bf A}(\bth)  = \begin{bmatrix}  
 -(k_{21} + k_{01}) & k_{12}                   & 0  \\
k_{21}                               & -(k_{12}+ k_{32}) & k_{23} \\
0     				& k_{32}                         & -k_{23}
\end{bmatrix} \; , 
 \qquad
{\bf C}(\bth) = \begin{bmatrix} 
c_{1} & 0 & 0  \\
0 & c_{2} & 0 \\
0 & 0 & c_{3} 
\end{bmatrix} \; ,
\end{gathered} \label{eq:parent_case}
\end{align}
where the parameter vector is
\begin{align*}
\bth = \begin{pmatrix} k_{01} & k_{12} & k_{21} & k_{23} & k_{32} & x_{0_1} & x_{0_2} & x_{0_3} & c_{1} & c_{2} & c_{3}  \end{pmatrix}^{\top} \in \bar{\mathbb{R}}_{+}^{11} \; .
\end{align*}

For simplicity, we have chosen to consider a parent structure that has a diagonal ${\bf C}$. By setting any $c_{i}=0$ ($i=1,2,3$), we readily produce an alternative structure (associated with an alternative experimental design) which models observations that are independent of $x_{i}$. 
For drawing the compartmental diagram associated with the parent structure or its variants (using procedure {\tt Uncontrolled\_Lin\_Comp\_Fig}, \ref{app:figure_code}), ${\bf A}$ directs us to record the parameters associated with flows out of the system with ${\tt outflow\_params} \triangleq \begin{bmatrix} k_{01}, &  0, &  0 \end{bmatrix}$.

\begin{figure}[!bt]
\centering
\includegraphics[scale=0.45]{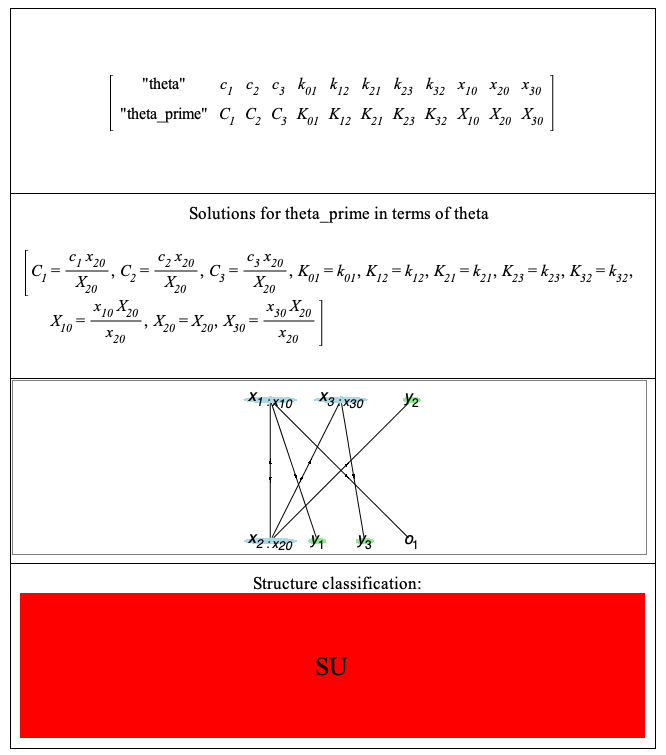}
\caption{Key features of the output window produced by application of Maple's {\tt Explore} to {\tt Explore\_SGI\_test} (Listing~\ref{code:Explore_SGI_routine}) in the study of our parent structure having representative system \eqref{eq:parent_case}.}
\label{fig:Explore_sshot}
\end{figure}

Figure~\ref{fig:Explore_sshot} shows the SGI test results (the result sections of the {\tt Explore} window) for the parent structure illustrated by \eqref{eq:parent_case}. The top panel shows $\bth$ and $\bth^{\prime}$ for ease of comparison. The third panel presents 
a compartmental diagram of the structure under consideration. The bottom panel shows the structure's classification. 

The second panel shows the solution set of the test equations. Here we see that some parameters are uniquely identifiable 
(e.g. $K_{01}=k_{01}$). Other parameters are free (e.g. $X_{20}=X_{20}$), leading to the structure's classification as SU. The solution also provides other insights. We note that we may rearrange the expression for $C_{1}$ to yield $C_{1}X_{20} = c_{1}x_{20}$. That is, whilst we cannot uniquely estimate $c_{1}$ and $x_{20}$ individually, we may be able to obtain a unique estimate of their product. This insight 
may guide the reparameterisation of the parent structure so as to remove one contributor to the structure's SU status. 

We can readily consider variants of the parent structure. Using the appropriate input box on the Explore dashboard, 
setting $c_{1}=1$ results in an SGI structure. Alternatively, modifying the parent structure by setting $c_{1}=c_{2}=1$ and $c_{3}=0$ yields an SLI structure.

\begin{remark}
Our procedures were designed for ULTI structures, however, we can also accommodate the experimental case where the initial condition of any state variable is set by an impulsive input, and there are no other applied inputs.
\end{remark}

\section{Concluding remarks} \label{s:conclusions}
We have presented Maple 2020 code to allow the interactive testing of a parent ULTI structure and its variants for SGI. 
Whilst we believe this to be a novel contribution, there are still opportunities to improve upon the presentation here. 
\begin{itemize}[noitemsep]
\item We used the workaround of an {\tt Array} so that {\tt Explore} could display multiple objects (not merely test results) in our interactive panel. This choice limited our control over object layout. Our presentation may be improved by designing an interactive application which uses ``embedded components''.
\item A diagram produced by {\tt Uncontrolled\_Lin\_Comp\_Fig} will be more informative if it could show each edge labelled with the appropriate parameter. At present, {\tt DrawGraph} is limited to showing numerical weights on edges. Hence, it will be useful to produce a new procedure (based on {\tt DrawGraph}) that does not have this restriction.
\end{itemize}
We also see opportunities to further the contributions of this chapter. An extension of {\tt Uncontrolled\_Lin\_Comp\_Fig} to suit controlled LTI structures will require modifications to include the influence of inputs on states. Certain complexities in the testing of controlled structures
(see \cite[Section~4]{Whyte_MATRIX_IDrev_2020}) will necessitate substantial changes to how our interactive application processes arguments. For example, it may be desirable to consider an SGI test where output is available for (the often realistic case of) a limited number of inputs that do not permit us to obtain the structure's invariants. The testing of structures of non-linear systems for SGI will require new methods for extracting invariants, and for displaying any edges which depend on state variables in a non-linear manner.

\subsubsection*{Acknowledgements} 
The author thanks the organisers of ``Identifiability problems in systems biology''  at the American Institute of Mathematics 
(San Jose, California, August 19--23, 2019) for the invitation to attend, and participants for useful discussions. This chapter's presentation benefited from the language definition for Maple code (for \LaTeX's listings package) by Maplesoft's Erik Postma. Appreciation also goes to an anonymous reviewer for helpful comments which informed or inspired various improvements to the original Maple code.

\appendix

\renewcommand{\thesection}{Appendix~\arabic{section}} 

\section{Maple code for drawing a modified compartmental diagram} \label{app:figure_code}
We use Listing~\ref{code:figure_code} in drawing a modified compartmental diagram of the model structure currently under investigation. When the {\tt Explore} window associated with Listing~\ref{code:Explore_application} is launched, the diagram displayed is updated in response to user selections from the drop-down ``layout'' menu or changes to the input boxes which set parameter values.

{ \scriptsize
\begin{lstlisting}[caption={Maple code which uses the definition of a model structure and some user-specified parameters drawn from this in drawing a modified compartmental diagram},label=code:figure_code, language=maple]
Uncontrolled_Lin_Comp_Fig:=proc(A::Matrix, C::Matrix, x0::Vector, x_colour::string, y_colour::string, outflow_params::list, tracing::truefalse, $)
local AugmentedA, Cmod, num_xy, zero_row_padding, AdjacencyA, AugA_rows, outflow_mat, outflow_detect, outflow_col, outflow_det_cols, AugA_cols, i, j, m, n, x0_list, xseq, xvertex_colouring, yvertex_colouring, vertex_colouring, num_outflows, outflow_labels, num_outrows, xlabels, ylabels, outflow_tag_indices, labels_list, G, outflow_indices, outflow_colour;  
description "Combine matrices A and C and initial state vector x0 of a defined linear time-invariant continuous-time structure to produce a diagram showing interconnections of state variables (x), initial conditions, outflows, and dependence of outputs (y) on x."; 
n:=LinearAlgebra[RowDimension](A);
if (type(C,Vector[row])=true) then Cmod:=convert(C,Matrix); 
else Cmod:=C; fi;
m:=LinearAlgebra[RowDimension](Cmod);
# As A is specified using the compartmental modelling convention, construct an adjacency matrix from transpose(A). This allows us to use the conventions of graph theory, and hence Maple's GraphTheory without modification.
AugmentedA:=LinearAlgebra[Transpose](A);
# Add to AugmentedA new columns showing dependence of y on x.
AugmentedA:=<AugmentedA | LinearAlgebra[Transpose](Cmod)>;
num_xy:=LinearAlgebra[ColumnDimension](AugmentedA);
zero_row_padding:=LinearAlgebra[ZeroMatrix](m, num_xy);
AugmentedA:=<AugmentedA, zero_row_padding>;
if (tracing=true) then print("AugmentedA with outputs", AugmentedA); fi;
# Note outflows by the presence of outflow_params in A's main diagonal.
if (outflow_params<>[]) then
AugA_rows:=LinearAlgebra[RowDimension](AugmentedA);
outflow_mat:=Matrix(AugA_rows,0);
for i from 1 to n do;
outflow_detect:={-outflow_params[i]} intersect {op(A[i,i])};
if (outflow_detect<>{}) then outflow_col:=Matrix(AugA_rows,1); outflow_col[i,1]:=op(outflow_detect); outflow_mat:=<outflow_mat | outflow_col>; fi; od; 
if (tracing=true) then print("outflow_mat",outflow_mat); fi;
# Append this to AugmentedA to add columns for the outflows; each gets a node, made invisible later.
AugmentedA:=<AugmentedA|outflow_mat>;
if (tracing=true) then print("AugA with outflows", AugmentedA); fi;
# Now add some more zero padding rows
outflow_det_cols:=LinearAlgebra[ColumnDimension](outflow_mat);
AugA_cols:=LinearAlgebra[ColumnDimension](AugmentedA);
zero_row_padding:=LinearAlgebra[ZeroMatrix](outflow_det_cols, AugA_cols);
AugmentedA:=<AugmentedA, zero_row_padding>;
if (tracing=true) then print("AugA fully padded", AugmentedA); fi;
fi; # end of block for processing outflows
AdjacencyA:=AugmentedA;
# Create an adjacency matrix for the graph, setting diagonal elements to zero to avoid self-loops
for i to LinearAlgebra[RowDimension](AdjacencyA) do;
    for j to LinearAlgebra[ColumnDimension](AdjacencyA) do;
if (AdjacencyA[i, j]<>0) then if (i=j) then AdjacencyA[i, j]:=0;
else AdjacencyA[i, j]:=1; fi; fi; 
od; od;
# Define the colouring of vertices by type
xvertex_colouring:=seq(x_colour, 1..n);
# We need to establish which outputs are active as a result of user inputs.
yvertex_colouring:=seq(y_colour, 1..m); 
vertex_colouring:=[xvertex_colouring, yvertex_colouring];
# Include any outflow with an "invisible" node.
num_outflows:=LinearAlgebra[ColumnDimension](outflow_mat);
if (num_outflows>0) then outflow_colour:="white";
vertex_colouring:=[op(vertex_colouring),seq(outflow_colour,1..num_outflows)]; 
outflow_labels:=outflow; fi;
G:=GraphTheory[Graph](AdjacencyA, vertexcolor=vertex_colouring); 
# Customise the vertex labels
xseq:=[seq(x[i], i=1..n)]; xseq:=map(convert,xseq,symbol); 
x0_list:=convert(x0,list); x0_list:=map(convert,x0_list,symbol);
xlabels:=convert(Vector[row](n),list);
for i from 1 to n do; xlabels[i]:=convert(StringTools[Join]([xseq[i],x0_list[i]]," : "),symbol); od;
if (tracing=true) then print("xlabels",xlabels); fi;
ylabels:=seq(y[i], i=1..m);
labels_list:=[op(xlabels), ylabels]; 
if (tracing=true) then print("labels_list",labels_list); fi;
if (num_outflows>0) then outflow_indices:=[`$`(num_xy+1..num_xy+num_outflows)];
outflow_tag_indices:=[seq(`if`(outflow_params[i] <> 0, i, NULL), i = 1..numelems(outflow_params))];
if (tracing=true) then print("outflow_tag_indices", outflow_tag_indices); fi;
labels_list:=[op(labels_list),seq(o[i],i=outflow_tag_indices)];  
if (tracing=true) then print("final labels_list", labels_list); fi;
fi;
G:=GraphTheory[RelabelVertices](G, labels_list); 
return G;
end proc
\end{lstlisting}
}

\section{Maple code to launch an Explore window} \label{app:Explore_launch}
Listing~\ref{code:Explore_application} presents the {\tt Explore} command which launches our interactive SGI test dashboard by invoking Listing~\ref{code:Explore_SGI_routine}. Here we consider the case of three state variables and three outputs; the user can readily change these details. To explain the parameters: ${\tt A}$ is the structure's ${\bf A(\bth)}$, {\tt p1}, {\tt p2}, {\tt p3} are the 
observation gain parameters on the leading diagonal of ${\bf C(\bth)}$, and {\tt p4}, {\tt p5}, {\tt p6} are the initial state parameters in ${\bf x_{0}}(\bth)$.
Initially, each of {\tt p1},\ldots,{\tt p6} are assigned a parameter symbol appropriate for their relationship to $\bth$. Each of
these six parameters may be changed through a text-input box.
Parameter {\tt p7} supplies a graph output style understood by {\tt DrawGraph}, initially (the widely applicable) ``default''. 
Output from other options (such as ``spring'') may be easier to interpret, but return an error
when any of {\tt p1}, {\tt p2}, or {\tt p3} are set to zero, causing the removal of a link between a state variable and its corresponding output.
Parameter {\tt p8} takes one of the two pre-defined values for {\tt theta\_mod\_type}, which dictates the method employed in creating {\tt theta\_prime} from {\tt theta} (used by {\tt theta\_prime\_creation}). The user changes {\tt p7} and {\tt p8} values by selecting an option from the relevant drop-down menu.
If logical-type parameter {\tt tracing:=true}, Maple will show the output of steps used in constructing the structure's compartmental diagram.

{ \scriptsize
\begin{lstlisting}[caption={Maple code using Maple's {\tt Explore} with {\tt Explore\_SGI\_routine} (Listing \ref{code:Explore_SGI_routine}) to produce an interactive panel.},label=code:Explore_application, language=maple]
Explore(Explore_SGI_test(A, [p1, p2, p3], [p4, p5, p6], outflow_params, p7, p8, s, canonical_form, tracing), parameters = [[p1, controller = textarea, label = "x1 observation gain",placement=left], [p2, controller = textarea, label = "x2 observation gain",placement=left], [p3, controller = textarea, label = "x3 observation gain",placement=left], [p4, controller = textarea, label = "x1 initial condition",placement=left], [p5, controller = textarea, label = "x2 initial condition",placement=left], [p6, controller = textarea, label = "x3 initial condition",placement=left], [p7=[default, bipartite, circle, planar, spectral, spring, tree], label=Layout, placement=left], [p8=["Caps","underscore"],label="theta_mod_type",placement=left]], initialvalues = [p1 = c__1, p2 = c__2, p3 = c__3, p4 = x__10, p5 = x__20, p6 = x__30, p7 = default], size = [650, 750], echoexpression = false, insert = true, title = "A use of 'Explore' to interactively test variants of a parent uncontrolled, linear time-invariant structure for structural global identifiability", overview = "Explore is used to generate the solutions for the test of a structure for SGI. Starting from a parent structure, by specifying values (or parameters in theta) we may change which states are observed, and which have non-zero initial conditions.");
\end{lstlisting}
}

\bibliographystyle{splncs04}
 \bibliography{/Users/whytej/OneDrive/Documents/Whyte_writingup_app/Bibliography_Moved_Jul8_2018/thesis, /Users/whytej/OneDrive/Documents/Whyte_writingup_app/Bibliography_Moved_Jul8_2018/identifiability_lss, /Users/whytej/OneDrive/Documents/Whyte_writingup_app/Bibliography_Moved_Jul8_2018/biapaper, /Users/whytej/OneDrive/Documents/Whyte_writingup_app/Bibliography_Moved_Jul8_2018/Structural_ID}

\end{document}